\documentclass[aps,pra,twocolumn,groupedaddress,nofootinbib]{revtex4-2}

\usepackage{amsmath,amssymb,bm,mathtools}
\usepackage{microtype}
\usepackage[hidelinks]{hyperref}
\usepackage{booktabs}

\newcommand{\kb}{k_{\mathrm B}}
\newcommand{\Tr}{\operatorname{Tr}}

\newcommand{\cF}{\mathcal{F}}

\begin{document}

\title{Comment on Statistical mechanics from quantum envariance and exchange symmetry}

\author{Ridha Horchani}
\email{horchani@squ.edu.om}
\affiliation{Department of Physics, College of Science, Sultan Qaboos University, Muscat, Oman}

\begin{abstract}
Ojha, Sardana, and Ghosh [Phys. Rev. A \textbf{113}, 042221 (2026)] propose that tracing environmental records of particle permutations produces an entropy $\kb\ln N!$, thereby explaining the Gibbs factor, and use the same construction to multiply the Saha equilibrium relation by $1/(N_e!N_p!)$. We show that these conclusions do not follow. The reduced density matrix printed in their Eq. (32) is not a partial trace, and even under the natural corrected interpretation the entropy equals $\kb\ln N!$ only when the system branch states are mutually orthogonal. That condition is not generally satisfied and, when imposed on labeled permutation branches, does not by itself restrict the particle state to one bosonic or fermionic symmetry sector. A two-particle calculation makes the contradiction explicit. Their Gibbs-paradox calculation also starts from a distinguishable-particle entropy while calling it the Sackur--Tetrode entropy. In the Saha section, the state in Eq. (52), as written, factorizes and has zero system--environment entanglement. The proposed factorial does not approach unity in the claimed dilute-gas limit, does not yield a finite nonzero intensive thermodynamic limit, and counts indistinguishability twice. We give the corrected canonical and fugacity-based formulations and identify which standard results of the paper remain unaffected.
\end{abstract}

\maketitle

\section{Introduction}

Reference~\cite{Ojha2026} seeks to formulate several equilibrium results of statistical mechanics using environment-assisted invariance. Its standard combinatorial limits leading from a binomial distribution to Poisson and Gaussian distributions are not at issue here. Likewise, once the usual bosonic or fermionic occupation-number constraints are imposed, the mode sums in Eqs.~(45)--(47) reproduce the familiar Bose--Einstein and Fermi--Dirac functions.

The central new claims are different. Reference~\cite{Ojha2026} argues that an environment that records particle permutations produces a von Neumann entropy
\begin{equation}
S_{\rm ent}=\kb\ln N!,
\label{eq:claimed_entropy}
\end{equation}
that supplies a microscopic explanation of the Gibbs factor, and it subsequently introduces
\begin{equation}
\Gamma_{\rm ind}^{\rm ent}=\frac{1}{N_e!N_p!}
\label{eq:claimed_gamma}
\end{equation}
as a new multiplicative correction to the Saha equation. These claims depend on Eqs.~(24)--(36) and (52)--(56) of Ref.~\cite{Ojha2026}. We show below that those equations contain mutually independent mathematical and thermodynamic inconsistencies.

The most direct problems are the following. First, Eqs.~(24)--(26) do not use the Sackur--Tetrode entropy, despite being presented as such. Second, Eq.~(32) is not the partial trace of Eq.~(31). Even after repairing its notation, the claimed entropy in Eq.~(33) is not generally obtained. Third, orthogonal environmental records destroy permutation coherence rather than project a state onto the bosonic or fermionic sector. Fourth, the positive von Neumann entropy of the resulting mixture is subsequently subtracted without a thermodynamic derivation. Fifth, the paper later treats $\ln N!$ as a logarithmic, sublinear quantity, in direct contradiction with Stirling's formula and with its own Eq.~(34). Finally, the Saha correction is based on a state that, as written, is a product state, double-counts Gibbs factors already present in the chemical potentials, depends on the arbitrary size of the chosen system, and cannot approach unity in the claimed dilute limit.

Equation numbers below refer to Ref.~\cite{Ojha2026} unless explicitly stated otherwise.

\section{The starting entropy in Eqs.~(24)--(26) is not the Sackur--Tetrode entropy}

For a nondegenerate monatomic ideal gas, the Sackur--Tetrode expression is
\begin{equation}
S_{\rm ST}(N,V,T)
= N\kb\left[\ln\left(\frac{V}{N\lambda_{\rm th}^3}\right)+\frac{5}{2}\right],
\label{eq:ST}
\end{equation}
where
\begin{equation}
\lambda_{\rm th}=\frac{h}{\sqrt{2\pi m\kb T}}.
\end{equation}
Consider two equal portions of the same gas, each containing $N$ particles in volume $V$ at the same temperature. Equation~\eqref{eq:ST} gives
\begin{align}
S_i &=2N\kb\left[\ln\left(\frac{V}{N\lambda_{\rm th}^3}\right)+\frac52\right],\\
S_f &=2N\kb\left[\ln\left(\frac{2V}{2N\lambda_{\rm th}^3}\right)+\frac52\right]=S_i.
\end{align}
Thus the Sackur--Tetrode formula gives
\begin{equation}
\Delta S_{\rm mix}=0
\end{equation}
for removal of a partition between identical gases of equal temperature and density.

By contrast, Eqs.~(24) and (25) of Ref.~\cite{Ojha2026} use an entropy of the form
\begin{equation}
S_{\rm d}=N\kb\left[\ln V+\frac32\ln T+\sigma\right],
\label{eq:dist_entropy}
\end{equation}
with $\sigma$ treated as independent of $N$. Equation~\eqref{eq:dist_entropy} lacks the $-N\kb\ln N$ contribution generated by $1/N!$ and is therefore the entropy associated with labeled, distinguishable-particle counting, up to constants. It is not the Sackur--Tetrode expression. The result $2N\kb\ln2$ in Eq.~(26) follows precisely because the $N$ dependence that removes the Gibbs paradox has already been omitted.

This point is also internally acknowledged later in Ref.~\cite{Ojha2026}, where the authors state that the Sackur--Tetrode formula already contains the $1/N!$ factor. Consequently, the sequence ``use Sackur--Tetrode $\rightarrow$ obtain a nonzero entropy of mixing $\rightarrow$ repair it'' is not a valid calculation. The correct statement is instead
\begin{equation}
\text{distinguishable counting} \quad\Longrightarrow\quad 2N\kb\ln2,
\end{equation}
whereas
\begin{equation}
\text{Sackur--Tetrode counting} \quad\Longrightarrow\quad 0.
\end{equation}
This correction is important because the later entanglement construction is presented as canceling an entropy that the correctly identified Sackur--Tetrode formula never produces.

\section{The partial trace and the entropy in Eqs.~(31)--(33)}

\subsection{Correct partial trace}

Write Eq.~(31) abstractly as
\begin{equation}
|\Psi\rangle_{SE}=\frac{1}{\sqrt{M}}
\sum_{\pi=1}^{M}g_\pi|\Phi_\pi\rangle_S|E_\pi\rangle_E,
\qquad M=N!,
\label{eq:global_general}
\end{equation}
where $|g_\pi|=1$ and the environment vectors satisfy
\begin{equation}
\langle E_\sigma|E_\pi\rangle=\delta_{\sigma\pi}.
\end{equation}
The reduced state is
\begin{align}
\rho_S
&=\Tr_E|\Psi\rangle\langle\Psi|\\
&=\frac{1}{M}\sum_{\pi,\sigma}g_\pi g_\sigma^*
|\Phi_\pi\rangle\langle\Phi_\sigma|
\langle E_\sigma|E_\pi\rangle\\
&=\frac{1}{M}\sum_{\pi}|\Phi_\pi\rangle\langle\Phi_\pi|.
\label{eq:correct_trace}
\end{align}
Equation~(32) of Ref.~\cite{Ojha2026}, however, as printed, is a sum of the same global projector $|\Psi_{SE}\rangle\langle\Psi_{SE}|$. Such a sum is simply that projector again and is not an operator on the system Hilbert space. If Eq.~(32) is interpreted literally, the state remains pure and Eq.~(33) does not follow. If Eq.~(32) is interpreted as an intended shorthand for Eq.~\eqref{eq:correct_trace}, the entropy problem remains, as shown next.

\subsection{The entropy equals $\kb\ln N!$ only in a special case}

The state in Eq.~\eqref{eq:correct_trace} is maximally mixed on an $M$-dimensional support only if the $M$ system vectors $|\Phi_\pi\rangle$ are mutually orthogonal. Orthogonality of the environment vectors is not sufficient. In general, the nonzero eigenvalues of $\rho_S$ coincide with those of the normalized Gram matrix
\begin{equation}
G_{\pi\sigma}=\frac{1}{M}\langle\Phi_\sigma|\Phi_\pi\rangle.
\end{equation}
Therefore,
\begin{equation}
S(\rho_S)\leq \kb\ln M=\kb\ln N!,
\label{eq:entropy_bound}
\end{equation}
with equality only if the branch states are orthonormal.

This is not a minor technical condition. For bosons, repeated occupations make many permutation branches identical. If occupations are $n_1,n_2,\ldots$, the number of distinct labeled permutations is at most
\begin{equation}
\frac{N!}{\prod_j n_j!},
\end{equation}
so even within a labeled tensor-product construction one cannot generally obtain $\kb\ln N!$. In the extreme case in which all bosons occupy one orbital, every permutation produces the same system vector and the entropy is zero.

\subsection{Explicit two-particle counterexamples}

Take two bosons in the same normalized one-particle state $|a\rangle$. The two permutation branches are identical,
\begin{equation}
|\Phi_{12}\rangle=|a\rangle_1|a\rangle_2
=|\Phi_{21}\rangle.
\end{equation}
Equation~\eqref{eq:global_general} then factorizes:
\begin{equation}
|\Psi\rangle_{SE}=|aa\rangle_S\otimes
\frac{|E_{12}\rangle+|E_{21}\rangle}{\sqrt2}.
\end{equation}
Hence
\begin{equation}
\rho_S=|aa\rangle\langle aa|,
\qquad S(\rho_S)=0,
\end{equation}
whereas Eq.~(33) predicts $\kb\ln2$. This single valid bosonic state disproves the claimed universal entropy.

A second example displays a deeper conflict. Let $|a\rangle$ and $|b\rangle$ be orthogonal and define
\begin{equation}
|ab\rangle=|a\rangle_1|b\rangle_2,
\qquad |ba\rangle=|b\rangle_1|a\rangle_2.
\end{equation}
Orthogonal environmental records give
\begin{equation}
\rho_S=\frac12\left(|ab\rangle\langle ab|+|ba\rangle\langle ba|\right).
\label{eq:two_mixture}
\end{equation}
Introduce the normalized exchange eigenstates
\begin{equation}
|\psi_\pm\rangle=\frac{|ab\rangle\pm|ba\rangle}{\sqrt2}.
\end{equation}
Then
\begin{equation}
\rho_S=\frac12|\psi_+\rangle\langle\psi_+|
+\frac12|\psi_-\rangle\langle\psi_-|.
\label{eq:wrong_sectors}
\end{equation}
For two bosons, the antisymmetric component is forbidden; for two fermions, the symmetric component is forbidden. When $E$ is an external environment, the reduced particle state in Eq.~\eqref{eq:wrong_sectors} is invariant under exchange as a density operator, but it is not supported solely in the bosonic or fermionic exchange sector required for the chosen species. Projection onto the appropriate sector gives
\begin{equation}
\rho_\pm=
\frac{P_\pm\rho_S P_\pm}{\Tr(P_\pm\rho_S)}
=|\psi_\pm\rangle\langle\psi_\pm|,
\end{equation}
which is pure and again has zero von Neumann entropy.

The construction therefore faces a dichotomy. If the permutation branches are physical symmetrized or antisymmetrized states, the reduced state is pure and $S(\rho_S)=0$. If instead the permutation branches are orthogonal labeled tensor products, the reduced particle state is not supported solely in the required bosonic or fermionic exchange sector. Neither alternative establishes Eq.~(33) as the entropy of an identical-particle gas.

\subsection{Orthogonal environmental records do not enforce indistinguishability}

An orthogonal record $|E_\pi\rangle$ makes the value of $\pi$ perfectly distinguishable in principle. Tracing over the record makes that information inaccessible to an observer restricted to $S$, but it does not identify the branches as one physical ray. Instead, it removes off-diagonal coherence and produces the incoherent mixture in Eq.~\eqref{eq:correct_trace}. This is the standard effect of which-alternative information.

By contrast, a bosonic or fermionic state belongs from the outset to the projected Hilbert space
\begin{equation}
\mathcal H_\pm=P_\pm\mathcal H^{\otimes N},
\qquad
P_\pm=\frac{1}{N!}\sum_{\pi\in S_N}(\pm1)^\pi U_\pi.
\label{eq:projector}
\end{equation}
The physical exchange requirement is
\begin{equation}
P_\pm\rho P_\pm=\rho,
\label{eq:sector_condition}
\end{equation}
not merely
\begin{equation}
U_\pi\rho U_\pi^\dagger=\rho.
\label{eq:commuting_condition}
\end{equation}
Equation~\eqref{eq:commuting_condition} is weaker and is satisfied by the mixture in Eq.~\eqref{eq:wrong_sectors}, which is not supported solely in the required bosonic or fermionic particle sector. This also affects the claim around Eqs.~(38)--(47): invariance of a reduced density operator under permutations does not by itself select the fully symmetric or fully antisymmetric representation. Those sectors must be supplied by the exchange postulate, as Ref.~\cite{Ojha2026} itself partly acknowledges.

\section{A positive reduced-state entropy cannot simply be subtracted to derive the Gibbs factor}

Equation~(36) defines
\begin{equation}
S_{\rm qm}=S_{\rm cl}-S_{\rm ent}.
\label{eq:manual_subtraction}
\end{equation}
No identity of quantum statistical mechanics gives Eq.~\eqref{eq:manual_subtraction} for the reduced state constructed in Eq.~(31). If a pure state becomes mixed after tracing an environment, its von Neumann entropy is the positive quantity $S(\rho_S)$. The trace operation does not simultaneously prescribe subtracting that entropy from an independently defined classical thermodynamic entropy.

The standard Gibbs correction has a different origin. In the Maxwell--Boltzmann regime, the canonical partition function is
\begin{equation}
Z_N^{\rm MB}=\frac{Z_1(\beta)^N}{N!},
\label{eq:MB_partition}
\end{equation}
which gives, relative to labeled counting,
\begin{equation}
\Delta S_{\rm counting}=-\kb\ln N!.
\label{eq:negative_counting}
\end{equation}
The sign in Eq.~\eqref{eq:negative_counting} is negative because the number of counted states is reduced. The state constructed using orthogonal environmental records instead has a positive mixing entropy. Reference~\cite{Ojha2026} obtains the desired sign only by placing a minus sign in Eq.~(36). Thus the numerical appearance of $\ln N!$ does not constitute a derivation of the Gibbs factor.

There is a second distinction. Equation~(28) is written as though
\begin{equation}
Z_N=\frac{Z_1^N}{N!}
\end{equation}
were the exact canonical partition function following from symmetrization or antisymmetrization. It is only the dilute Maxwell--Boltzmann limit. The exact ideal-gas expression is
\begin{equation}
Z_N^{\pm}=\frac{1}{N!}
\sum_{\pi\in S_N}(\pm1)^\pi
\Tr\left(U_\pi e^{-\beta H_N}\right),
\label{eq:exact_projected_Z}
\end{equation}
where nonidentity permutations generate exchange corrections. For $N=2$,
\begin{equation}
Z_2^{\pm}=\frac12\left[Z_1(\beta)^2\pm Z_1(2\beta)\right].
\label{eq:Z2}
\end{equation}
Only when $Z_1(2\beta)/Z_1(\beta)^2$ is negligible does Eq.~\eqref{eq:Z2} reduce to $Z_1^2/2!$. Hence the Gibbs factor and quantum exchange corrections are not the same object. The $1/N!$ factor remains necessary in the classical dilute gas, while the additional permutation-cycle terms vanish with the phase-space density. Statements in Ref.~\cite{Ojha2026} that the $1/N!$ term itself becomes important only near degeneracy conflate these two roles.

\section{The stated scaling of $\ln N!$ is mathematically incorrect}

After correctly writing in Eq.~(34) that
\begin{equation}
\ln N!\simeq N\ln N-N,
\end{equation}
Ref.~\cite{Ojha2026} states below Eq.~(54) that $\ln N!$ is weakly or sublinearly dependent on system size, scales as $\ln N$, and gives a per-particle contribution that vanishes in the thermodynamic limit. Each statement is false. Stirling's expansion is
\begin{equation}
\ln N!=N\ln N-N+\frac12\ln(2\pi N)+O(N^{-1}),
\label{eq:stirling}
\end{equation}
so
\begin{equation}
\frac{\ln N!}{N}=\ln N-1+O\left(\frac{\ln N}{N}\right).
\label{eq:per_particle}
\end{equation}
The quantity in Eq.~\eqref{eq:per_particle} does not tend to zero; it diverges logarithmically.

The familiar ideal-gas free energy remains extensive because the $N\ln N$ term from $\ln N!$ combines with the $N\ln V$ term from $Z_1^N$ at fixed density. An isolated additive term $\kb T\ln N!$ is not a small nonextensive offset. This distinction becomes decisive in the proposed Saha correction, where the factorials are introduced without the compensating volume dependence required for a finite thermodynamic limit.

\section{The proposed Saha correction}

\subsection{Equation~(52), as written, is a product state}

Equation~(52) has the form
\begin{equation}
|\Psi\rangle_{SE}=\frac{1}{\sqrt{N_e!N_p!}}
\sum_\pi |H,p,e\rangle_S\otimes|E_\pi\rangle_E.
\label{eq:saha_state}
\end{equation}
The system ket in Eq.~\eqref{eq:saha_state} carries no $\pi$ label. Therefore,
\begin{equation}
|\Psi\rangle_{SE}=|H,p,e\rangle_S\otimes
\left(\frac{1}{\sqrt{N_e!N_p!}}\sum_\pi|E_\pi\rangle_E\right).
\end{equation}
It is a product state. Its reduced state and entropy are
\begin{equation}
\rho_S=|H,p,e\rangle\langle H,p,e|,
\qquad S(\rho_S)=0.
\end{equation}
Thus Eqs.~(53) and (54) do not follow from Eq.~(52). If a missing permutation label on the system ket is supplied, the construction returns to the problems already demonstrated for Eqs.~(31)--(33).

Moreover, under the paper's proposed mechanism, permutations of the identical neutral-hydrogen atoms would also require treatment. Their omission is unexplained if factorial corrections are applied to electrons and protons on the basis of identical-particle permutations.

\subsection{The factorial does not approach unity in the claimed dilute limit}

The paper defines
\begin{equation}
\Gamma_{\rm ind}^{\rm ent}=\frac{1}{N_e!N_p!},
\qquad N_i=n_iV,
\end{equation}
and asserts that $\Gamma_{\rm ind}^{\rm ent}\to1$ when $n_i\lambda_i^3\ll1$. The definition contains neither $\lambda_i$ nor $T$, so this conclusion cannot follow.

At fixed particle numbers, taking the dilute limit by increasing $V$ leaves $\Gamma_{\rm ind}^{\rm ent}$ unchanged. For example,
\begin{equation}
N_e=N_p=2
\quad\Longrightarrow\quad
\Gamma_{\rm ind}^{\rm ent}=\frac14
\end{equation}
at every density and temperature. At fixed nonzero densities, the thermodynamic limit gives $N_i\to\infty$ and hence $\Gamma_{\rm ind}^{\rm ent}\to0$, including in an arbitrarily dilute Maxwell--Boltzmann gas. The stated crossover between Eqs.~(51) and (56) is therefore absent.

What does vanish when $n\lambda^3\to0$ is the exchange-cycle contribution relative to the identity-permutation term in Eq.~\eqref{eq:exact_projected_Z}; the Gibbs factorial itself does not approach unity.

\subsection{No acceptable intensive thermodynamic limit and dependence on the arbitrary observation volume}

At fixed densities, the proposed correction depends on the arbitrary volume used to describe the same homogeneous plasma:
\begin{equation}
\Gamma(V)=\frac{1}{(n_eV)!(n_pV)!}.
\end{equation}
Replacing $V$ by $2V$ changes the right-hand side of the proposed Saha relation by
\begin{equation}
\frac{\Gamma(2V)}{\Gamma(V)}
=\frac{(N_e!)(N_p!)}{(2N_e)!(2N_p)!},
\end{equation}
although all intensive properties are unchanged.

The corresponding free-energy shift proposed in Eq.~(55) behaves, using Eq.~\eqref{eq:stirling}, as
\begin{align}
\frac{\Delta F}{V}
&=\frac{\kb T}{V}\left[\ln N_e!+\ln N_p!\right]\\
&\simeq \kb T\left[n_e\ln(n_eV)+n_p\ln(n_pV)-n_e-n_p\right].
\end{align}
It diverges as $(n_e+n_p)\kb T\ln V$ and therefore does not possess a finite, nonzero intensive thermodynamic limit. An equilibrium relation among densities cannot depend factorially on the arbitrary amount of material included in the bookkeeping volume.

There is also a sign inconsistency between Eqs.~(54) and (55). Equation~(54) defines the positive quantity
\begin{equation}
S_{\rm ent}=-\kb\ln\Gamma_{\rm ind}^{\rm ent}.
\end{equation}
Literal substitution into $-T S_{\rm ent}$ gives $+\kb T\ln\Gamma_{\rm ind}^{\rm ent}$, whereas Eq.~(55) prints $-\kb T\ln\Gamma_{\rm ind}^{\rm ent}$. The latter sign would correspond to the negative physical counting correction $\Delta S=\kb\ln\Gamma=-S_{\rm ent}$, not to the positive entropy defined in Eq.~(54). The notation therefore switches between an entanglement entropy and its negative.

\subsection{Indistinguishability is counted twice}

Equation~(49) is explicitly derived from the canonical partition function of an indistinguishable ideal gas:
\begin{equation}
\mu_i=\epsilon_i^0+\kb T\ln\left(\frac{n_i\lambda_i^3}{g_i}\right),
\label{eq:classical_mu}
\end{equation}
where an optional species ground energy $\epsilon_i^0$ has been written explicitly. The factorials $1/N_i!$ are exactly what produce the $n_i$ dependence in Eq.~\eqref{eq:classical_mu}. Applying chemical equilibrium,
\begin{equation}
\mu_H=\mu_p+\mu_e,
\end{equation}
then gives
\begin{equation}
\frac{n_pn_e}{n_H}
=\frac{g_pg_e}{g_H}
\frac{\lambda_H^3}{\lambda_p^3\lambda_e^3}
\exp\left(-\frac{E_I}{\kb T}\right),
\label{eq:standard_saha}
\end{equation}
which is Eq.~(50). The Gibbs factors have already been used. Multiplying Eq.~\eqref{eq:standard_saha} by $1/(N_e!N_p!)$ applies them a second time and is the source of the nonintensive result.

The same conclusion follows directly from the canonical partition function of the reacting mixture,
\begin{equation}
\mathcal Z(N_H,N_p,N_e)
=\prod_{i=H,p,e}\frac{q_i^{N_i}}{N_i!},
\qquad
q_i=\frac{g_iV}{\lambda_i^3}e^{-\beta\epsilon_i^0}.
\label{eq:mixture_partition}
\end{equation}
For one ionization event, the ratio of neighboring canonical weights is
\begin{equation}
\frac{\mathcal Z(N_H-1,N_p+1,N_e+1)}
{\mathcal Z(N_H,N_p,N_e)}
=\frac{N_H}{(N_p+1)(N_e+1)}\frac{q_pq_e}{q_H}.
\label{eq:neighbor_ratio}
\end{equation}
At the maximum of the distribution and in the thermodynamic limit, Eq.~\eqref{eq:neighbor_ratio} yields Eq.~\eqref{eq:standard_saha}. There is no remaining global factorial multiplier.

\subsection{Correct quantum-degenerate generalization}

If ideal-gas quantum degeneracy is to be retained, the correction is expressed through fugacities, not through factorials of total particle numbers. Define
\begin{equation}
z_i=\exp\left[\beta(\mu_i-\epsilon_i^0)\right].
\end{equation}
For an ideal Bose or Fermi species,
\begin{equation}
n_i=\frac{g_i}{\lambda_i^3}\cF_{3/2}^{(i)}(z_i),
\label{eq:quantum_density}
\end{equation}
where
\begin{equation}
\cF_{3/2}^{(B)}(z)=\operatorname{Li}_{3/2}(z),
\qquad
\cF_{3/2}^{(F)}(z)=-\operatorname{Li}_{3/2}(-z).
\end{equation}
Chemical equilibrium gives the intensive relation
\begin{equation}
\frac{z_pz_e}{z_H}=e^{-\beta E_I}.
\label{eq:fugacity_saha}
\end{equation}
Equations~\eqref{eq:quantum_density} and \eqref{eq:fugacity_saha} are the ideal quantum-statistical replacement of the classical Saha equation. Since $\cF_{3/2}(z)=z+O(z^2)$, the standard Saha result is recovered smoothly when all fugacities are small. Corrections are controlled by $n_i\lambda_i^3$, are intensive, and possess a thermodynamic limit. In a dense real plasma, interaction effects, screening, pressure ionization, and continuum lowering must additionally be treated; none is represented by Eq.~\eqref{eq:claimed_gamma}.

\section{Dimensional inconsistency in Eqs.~(40)--(43)}

Reference~\cite{Ojha2026} writes $\Omega_B=e^{S_B/\kb}$ and then expands the dimensional entropy as
\begin{equation}
S_B(E-E_S,N-N_S)\simeq S_B(E,N)-\beta E_S+\beta\mu N_S,
\end{equation}
while defining $\beta$ through derivatives of $S_B$. This mixes two conventions. For a dimensional entropy, the correct expansion is
\begin{equation}
S_B(E-E_S,N-N_S)
\simeq S_B(E,N)-\frac{E_S}{T}+\frac{\mu N_S}{T}.
\label{eq:dimensional_expansion}
\end{equation}
Equivalently, for the dimensionless entropy $s_B=S_B/\kb$,
\begin{align}
s_B(E-E_S,N-N_S)
&\simeq s_B(E,N)-\beta E_S+\beta\mu N_S,\\
\beta&=\frac{1}{\kb T}.
\label{eq:dimensionless_expansion}
\end{align}
Equations~(40)--(43) become dimensionally consistent only after choosing one of Eqs.~\eqref{eq:dimensional_expansion} or \eqref{eq:dimensionless_expansion}. The final grand-canonical exponential is standard, but it is not obtained from the definitions as printed without restoring the missing factors of $\kb$.

\section{Consequences for the claims of Ref.~\cite{Ojha2026}}

The errors above affect the main original conclusions as follows.

First, the environment-record construction does not derive the Gibbs factor. Equation~(33) fails for valid bosonic states, and when it does produce $\kb\ln N!$ it produces a mixture of labeled permutation branches rather than a state restricted to the physical bosonic or fermionic sector. The subsequent subtraction in Eq.~(36) inserts the desired sign rather than deriving it.

Second, the treatment does not show that indistinguishability is dynamically enforced by environmental entanglement. Orthogonal environmental records encode which-permutation information and decohere the corresponding branches. Particle indistinguishability is imposed by restricting observables and states to the appropriate exchange sector, represented by Eq.~\eqref{eq:projector}, independently of whether an external environment records anything.

Third, the proposed modified Saha equation is not a valid intensive thermodynamic relation. Its additional factor follows neither from the state in Eq.~(52) nor from the partition function, does not approach unity in the claimed dilute-gas limit, and does not yield a finite, nonzero intensive thermodynamic limit. It also leaves the treatment of identical neutral-hydrogen atoms unexplained and double-counts indistinguishability already included in Eq.~(49). Accordingly, the claimed suppression of ionization in dense plasmas and recombination-era cosmology cannot be inferred from Eq.~(56).

These conclusions do not invalidate the standard formulas reproduced elsewhere in Ref.~\cite{Ojha2026}. The binomial, Poisson, and Gaussian relations remain standard consequences of the assumed fine-grained weights and combinatorial limits. Equations~(45)--(47) remain the standard Bose--Einstein and Fermi--Dirac mode sums once bosonic or fermionic occupation constraints are supplied. What fails is the proposed identification of the Gibbs factor and the Saha modification with entanglement generated by environmental permutation records.

\section{Conclusion}

A consistent formulation separates three distinct notions that are conflated in Ref.~\cite{Ojha2026}: the Gibbs factor in Maxwell--Boltzmann counting, exchange-cycle corrections in degenerate quantum gases, and von Neumann entropy generated by tracing an environment. The Gibbs factor is implemented through the physical state space or through the projected partition function. Quantum exchange corrections are the nonidentity-permutation contributions in Eq.~\eqref{eq:exact_projected_Z} and vanish in the dilute limit. Environmental decoherence can generate a positive reduced-state entropy, but that entropy is neither universally $\kb\ln N!$ nor a quantity that may be subtracted from a classical entropy without an independent thermodynamic argument.

The corrected treatment gives the ordinary Sackur--Tetrode and Saha equations in the nondegenerate regime. For a degenerate ideal mixture, the appropriate generalization is the fugacity system in Eqs.~\eqref{eq:quantum_density} and \eqref{eq:fugacity_saha}. The factorial multiplier in Eq.~(56) is therefore not supported by the displayed derivation and should be removed. The claims based on the resulting ionization suppression should be corrected or replaced by a calculation using intensive quantum-statistical chemical potentials.

\end{document}